\documentclass[onecolumn,showpacs,12pt]{revtex4}
\usepackage{graphicx}
\usepackage{dcolumn}
\usepackage{bm}
\begin{document}
%\preprint{APS/123-QED}
%%%%%%%%%%%%%%%%%%%%%%%%
\newcommand{\hs}{\hspace*{0.5cm}}
\newcommand{\vs}{\vspace*{0.5cm}}
\newcommand{\be}{\begin{equation}}
\newcommand{\ee}{\end{equation}}
\newcommand{\bea}{\begin{eqnarray}}
\newcommand{\eea}{\end{eqnarray}}
\newcommand{\ben}{\begin{enumerate}}
\newcommand{\een}{\end{enumerate}}
\newcommand{\bde}{\begin{widetext}}
\newcommand{\ede}{\end{widetext}}
\newcommand{\nn}{\nonumber}
\newcommand{\crn}{\nonumber \\}
\newcommand{\Tr}{\mathrm{Tr}}
\newcommand{\non}{\nonumber}
\newcommand{\noi}{\noindent}
\newcommand{\al}{\alpha}
\newcommand{\la}{\lambda}
\newcommand{\bet}{\beta}
\newcommand{\ga}{\gamma}
\newcommand{\va}{\varphi}
\newcommand{\om}{\omega}
\newcommand{\pa}{\partial}
\newcommand{\+}{\dagger}
\newcommand{\fr}{\frac}
\newcommand{\bc}{\begin{center}}
\newcommand{\ec}{\end{center}}
\newcommand{\Ga}{\Gamma}
\newcommand{\de}{\delta}
\newcommand{\De}{\Delta}
\newcommand{\ep}{\epsilon}
\newcommand{\varep}{\varepsilon}
\newcommand{\ka}{\kappa}
\newcommand{\La}{\Lambda}
\newcommand{\si}{\sigma}
\newcommand{\Si}{\Sigma}
\newcommand{\ta}{\tau}
\newcommand{\up}{\upsilon}
\newcommand{\Up}{\Upsilon}
\newcommand{\ze}{\zeta}
\newcommand{\ps}{\psi}
\newcommand{\Ps}{\Psi}
\newcommand{\ph}{\phi}
\newcommand{\vph}{\varphi}
\newcommand{\Ph}{\Phi}
\newcommand{\Om}{\Omega}
%%%%%%%%%%%%%%%%%%%%%%%%

\title{A possible minimal gauge-Higgs unification}

\author{P. V. Dong}
\email {pvdong@iop.vast.ac.vn} \affiliation{Institute of Physics,
VAST, 10 Dao Tan, Ba Dinh, Hanoi, Vietnam\footnote{Permanent address}}
\affiliation{PH-TH Division, CERN, CH-1211 Geneva 23, Switzerland}

\date{\today}

\begin{abstract}

A possible minimal model of the gauge-Higgs  unification based on
the higher dimensional spacetime $M^4\otimes (S^1/Z_2)$ and the
bulk gauge symmetry $SU(3)_C\otimes SU(3)_W \otimes U(1)_X$ is
constructed in some details. We argue that the Weinberg angle and
the electromagnetic current can be correctly identified if one
introduces the extra $U(1)_X$ above and a bulk scalar triplet. The
VEV of this scalar as well as the orbifold boundary conditions
will break the bulk gauge symmetry down to that of the standard
model. A new neutral zero-mode gauge boson $Z'$ exists that gains
mass via this VEV. We propose a simple fermion content that is
free from all the anomalies when the extra brane-localized chiral
fermions are taken into account as well. The issues on recovering
a standard model chiral-fermion spectrum with the masses and
flavor mixing are also discussed, where we need to introduce the
two other brane scalars which also contribute to the $Z'$ mass in
the similar way as the scalar triplet. The neutrinos can get small
masses via a type~I seesaw mechanism. In this model, the mass of
the $Z'$ boson and the compactification scale are very constrained
as respectively given in the ranges: $2.7\ \mathrm{TeV}<
m_{Z'}<13.6\ \mathrm{TeV}$ and $40\ \mathrm{TeV}< 1/R < 200\
\mathrm{TeV}$.

\end{abstract}

\pacs{11.10.Kk, 12.60.-i}

\maketitle

\section{\label{intro}Introduction}

The standard model of fundamental particles  and interactions has
been very successful in describing observed phenonmena. However, a
serious problem exists if it tries to match a certain new physics
at much higher energy scales. In deed, the squared-mass parameter
of the Higgs field in the standard model receives quadratically
divergent radiative corrections. These divergences imply that the
low-energy parameter is sensitive to contributions of heavy fields
with masses lying at the cut-off scale, which in principle can
reach the Planck scale. The physics at the weak scale is strongly
disturbed that requires a striking cancellation between the
various contributions and/nor the bare parameter. However, this
introduces an unjustifiably huge fine-tuning from the Planck scale
down to the weak scale of the order
$M_{\mathrm{Pl}}/M_{\mathrm{weak}}\sim 10^{16}$, known as the
hierarchy problem. The scalar sector of the standard model is thus
not natural. This indicates that there should be underlining
principles that prevent finite Higgs masses from the radiative
corrections. A typical example is supersymmetry, where such
divergences can be removed by a symmetry relating boson and
fermion \cite{martin}. Another example is the Randall-Sundrum
model where the hierarchy can be understood via a warped factor
associated with the bulk dimension \cite{randallsundrum}. For
other approaches, let us call the reader's attention to Refs.
\cite{add-lh-tc}.

An alternative to solve the hierarchy  problem is the gauge-Higgs
unification \cite{ghu}, which is recently called for much
attention
\cite{hierarchy-ghu,gaugeredu,overlap-mixing,others,wothers}. In
this scenario, the Higgs fields are identified as extra
dimensional components of the gauge fields, that results in a bulk
gauge vector transforming as adjoint representations under a
higher dimensional  gauge symmetry. The gauge symmetry will
protect a finite Higgs mass against radiative corrections. In
fact, the gauge symmetry actually preventing the Higgs fields from
obtaining a higher dimensional mass is spontaneously broken by the
compactification; As a result, a finite effective mass term is
allowed and gets naturally generated as the sum of all radiative
corrections necessarily independent of the cut-off scale
\cite{hierarchy-ghu}. A nice feature of the theory is that the
Yukawa interactions become universal with an unique coupling
constant as combined with the gauge interactions in higher
dimensional spacetime. The low energy chiral fermions and residual
gauge symmetry can be recognized if the theory is compactified on
orbifolds, for example, $S^1/Z_2$ (which will be used throughout
the work) \cite{chiralfermion,gaugeredu}. The hierarchical
smallness of the Yukawa couplings at the low energy may arise from
wave-function overlap profiles associated with extra space. And, a
consistent fermion content with flavor mixing may be original from
couplings with heavy brane-localized chiral fermions
\cite{overlap-mixing}.

Because the Higgs and gauge bosons of the standard model lie in
the adjoint representations, the smallest gauge symmetry of which
must be $SU(3)_C\otimes SU(3)_W$. This simplest version has been
extensively studied since the birth of the gauge-Higgs unification
until now. The problems with it are that (i) the Weinberg angle
$s^2_W=3/4$ is not correct, (ii) because the generators of
non-Abelian Lie group $SU(3)_W$ is normalized such as
$\mathrm{Tr}[T_aT_b]=\de_{ab} /2$, the electromagnetic
interactions are wrong. For examples, the electromagnetic coupling
of the neutrinos is non-zero that equals to that of up quarks;
similarly the coupling of electron, muon or tau is the same as
that of down quarks. In this work we will provide a solution to
these problems by adding a group factor $U(1)_X$ to the gauge
symmetry and imposing a bulk scalar triplet. The nature of this
scalar as well as introductory of the other brane scalars will be
discussed accordingly in the text.

In the literature, the bulk fermions  have usually been assigned
with large rank $SU(3)_W$ representations such as sextets, octets,
and even 10-plets. In this work we propose a simpler fermion
content which contains only bulk triplets or antitriplets. The
heavy brane-localized chiral fermions are also introduced to make
unwanted fermions heavy and producing flavor mixing. The presence
of these exotic chiral fermions have another effect that all the
chiral anomalies at branes are necessarily cancelled out
\cite{anomaly}.

The rest of this work is organized as follows. In Sec.
\ref{model}, we introduce the model with stressing on the gauge
symmetry, orbifold boundary conditions, zero mode fields, proposal
of bulk scalar triplet, and symmetry breakings. In Sec.
\ref{gabo}, we diagonalize the mass matrix of zero mode neutral
gauge bosons, identifying physical fields and matching of gauge
coupling constants and Weinberg angle. Constraints on the new
physics are also given. Section \ref{fect} is devoted to a fermion
content of the model, presenting the way to identify the standard
model fermions, mass generations, and flavor mixings. The small
masses of neutrinos are also obtained. Section \ref{anoeff}
presents anomaly cancellations. Finally, we summarize our results,
and make conclusions and some remarks on effective potential in
the last section--Sec. \ref{concl}.

\section{\label{model}The model}

The five dimensional (5D) spacetime is  supposed to be a direct
product of the ordinary four dimensional (4D) Minkowski spacetime
$M^4$ and an orbifold $S^1/Z_2$ with a radius $R$ of $S^1$, namely
$M^4 \otimes (S^1/Z_2)$. The 5D coordinate is generally denoted as
$x^M=(x^\mu,y)$ in which $\mu=0,1,2,3$ and $y=x^5$. The symmetry
of the orbifold, namely the symmetries on the fifth dimension due
to $S^1: y\rightarrow y+2\pi R$ and $Z_2: y\rightarrow -y$,
implies the following two basic transformations $Z_i:
y_i+y\rightarrow y_i-y$ around the orbifold fixed points $y_i$
$(i=0,\pi,\ \mathrm{and}\ y_{0}=0,\ y_{\pi}=\pi R)$, also called
boundaries or branes. The orbifold is therefore exact an interval
$y\in[0,\pi R]$ of the length $\pi R$.

The 5D gauge symmetry is considered as  $SU(3)_C \otimes SU(3)_W
\otimes U(1)_X$. The $SU(3)_W$ is the most minimal group that
contains electroweak gauge bosons and a Higgs doublet. The
$U(1)_X$ is needed to recover correct Weinberg angle and
electromagnetic current. Assuming $G_M, g_s\in SU(3)_C$, $A_M,
g\in SU(3)_W$ and $B_M, g_x \in U(1)_X$ as their corresponding 5D
gauge bosons and gauge coupling constants, and $\psi$ as a general
5D fermion multiplet, we have the following Lagrangian (up to the
gauge fixing and ghost terms): \bea \mathcal{L}&=&-\fr 1 2
\mathrm{Tr} G_{MN} G^{MN}-\fr 1 2 \mathrm{Tr} A_{MN} A^{MN}-\fr 1
4 B_{MN} B^{MN}\crn && +\bar{\psi}(iD\!\!\!\!/ -M_\psi
\epsilon(y))\psi,\eea where
\bea B_{MN}&=&\pa_M B_N-\pa_N B_M, \\
A_{MN}&=&\pa_M A_N - \pa_N A_M + ig [A_M,A_N],\\
G_{MN}&=&\pa_M G_N - \pa_N G_M + ig_s [G_M,G_N],\\
D\!\!\!\!/ &=& \Ga^M (\pa_M + ig_s G_M + ig A_M +ig_x X B_M), \eea
where $X$ is the charge of $U(1)_X$, $A_M\equiv T_a A^a_M$ with
$T_a$ ($a=1,2,...,8$) being the generators of $SU(3)_W$ and
satisfying $\mathrm{Tr}[T_aT_b]=\fr 1 2 \de_{ab}$, similarly for
$G_M$, $\Ga^M\equiv (\gamma^\mu,i\gamma^5)$ so that
$\left\{\Ga^M,\Ga^N\right\}=2g^{MN}=2\mathrm{diag}(1,-1,-1,-1,-1)$
defining a Clifford algebra of the 5D spacetime symmetry,
$\epsilon(y)=|y|/y$ is the sign function of $y$, and $M_\psi$ is
the bulk kink mass term for $\psi$.

Under the orbifold symmetries, the Lagrangian is invariant  and we
have boundary conditions on the fields as follows: \bea
\left[\begin{array}{c}
         G_\mu (x,y_i-y)  \\
         G_5 (x,y_i-y)  \\
      \end{array}\right]
       =
       P_i\left[\begin{array}{c}
         G_\mu(x,y_i+y)  \\
         - G_5(x,y_i+y)  \\
      \end{array}\right] P^{-1}_i,
     \eea
where $P_i$ is the representation of $Z_i$ (which including  all
the following similar ones have properties  unitary and
hermitian), acting on 5D gluons and thus identified as
$P_i=\mathrm{diag}(1,1,1)$. In terms of parity-value pair $(P_0\
P_\pi)$, we can explicitly write \bea  G_\mu =
\left[\begin{array}{ccc}
           (+\ +) & (+\ +) & (+\ +)\\
           (+\ +) & (+\ +) & (+\ +)\\
           (+\ +) & (+\ +) & (+\ +)\\
\end{array}\right],\hs\hs  G_5= \left[\begin{array}{ccc}
           (-\ -) & (-\ -) & (-\ -)\\
           (-\ -) & (-\ -) & (-\ -)\\
           (-\ -) & (-\ -) & (-\ -)\\
\end{array}\right],\ \eea which imply that the zero modes of $G_\mu$ are exact
ordinary 4D gluons of the standard model, while $G_5$ does not
have any zero mode.

For electroweak gauge bosons, we have also \bea
\left[\begin{array}{c}
         A_\mu (x,y_i-y)  \\
         A_5 (x,y_i-y)  \\
      \end{array}\right]
       &=&
       P_i\left[\begin{array}{c}
         A_\mu(x,y_i+y)  \\
         - A_5(x,y_i+y)  \\
      \end{array}\right] P^{-1}_i,\\
      \left[\begin{array}{c}
         B_\mu (x,y_i-y)  \\
         B_5 (x,y_i-y)  \\
      \end{array}\right]
       &=&
     \left[\begin{array}{c}
         B_\mu(x,y_i+y)  \\
         - B_5(x,y_i+y)  \\
      \end{array}\right],
\eea where $P_i$ is another representation of $Z_i$ that acts on
electroweak bosons, and chosen as $P_i=\mathrm{diag}(-1,-1,1)$.
Explicitly we write \bea  A_\mu = \left[\begin{array}{ccc}
           (+\ +) & (+\ +) & (-\ -)\\
           (+\ +) & (+\ +) & (-\ -)\\
           (-\ -) & (-\ -) & (+\ +)\\
        \end{array}\right],\hs\hs  A_5= \left[\begin{array}{ccc}
           (-\ -) & (-\ -) & (+\ +)\\
           (-\ -) & (-\ -) & (+\ +)\\
           (+\ +) & (+\ +) & (-\ -)\\
        \end{array}\right]. \eea
The zero modes of $A_\mu$ and $B_\mu$ will contain the standard
model electroweak gauge bosons and a new neutral gauge boson, i.e.
$A^1_\mu,\ A^2_\mu\sim W$ and $A^3_\mu,\ A^8_\mu,\ B_\mu \sim
\gamma,\ Z,\ Z'$ (new). The zero modes of $A_5$ will contain the
standard model Higgs doublet: $(A^4_5-iA^5_5,\ A^6_5-iA^7_5)^T\sim
H$. Here all the superscripts are the indices of $SU(3)_W$ adjoint
representation. The remaining gauge bosons including $B_5$ do not
have zero mode. They as well as $G_5$ and all the Kaluza-Klein
(KK) excitations, which have masses equal to or larger than
compactification or KK scale $1/R$, must be typically heavy.

The fermion $\psi$, that may be a triplet or an antitriplet  of
$SU(3)_W$, will satisfy the following boundary condition: \be
\psi(x,y_i-y) = P_i \ga^5 \psi(x,y_i+y).\ee Writing $\psi=\psi_L +
\psi_R$, the components will have simpler transformational rules:
\bea \psi(x,y_i-y)_L = - P_i \psi(x,y_i+y)_L,\hs\hs
\psi(x,y_i-y)_R = P_i \psi(x,y_i+y)_R,\eea which yield \be \psi_L
=   \left[\begin{array}{c}
            (+\ +)  \\
           (+\ +) \\
           (-\ -)\\
\end{array}\right]_L,\hs \hs \psi_R =   \left[\begin{array}{c}
            (-\ -)  \\
           (-\ -) \\
           (+\ +)\\
\end{array}\right]_R.\ee The zero modes are just a left-handed doublet or antidoublet and
a right-handed singlet under the standard model symmetry,
responsible for ordinary leptons and quarks. All other fermions
including the KK excitations must be heavy. Details on the fermion
content of the model will be provided in the next sections.

The expansions in KK modes for 5D fields so that their boundary
conditions are satisfied have been done. For a general gauge field
$V_{\mu,5}$, putting $M_n=n/R$ we have \bea
V_{\mu,5}(x,y)&=&\fr{1}{\sqrt{2\pi R}}V_{\mu,5}(x)
+\fr{1}{\sqrt{\pi R}} \sum^\infty_{n=1}
V^{(n)}_{\mu,5}(x)\cos(M_n y)\hs    (\mathrm{even}),\\
V_{\mu,5}(x,y)&=&\fr{1}{\sqrt{\pi R}} \sum^\infty_{n=1}
V^{(n)}_{\mu,5}(x)\sin(M_n y)\hs (\mathrm{odd}).\eea The mode
expansion for the fermion is quite different from those of gauge
fields due to the presence of the bulk kink mass term: \bea
\psi(x,y)=\left[
\begin{array}{c}
           \psi_L(x) f_L (y)+\sum_{n=1}^\infty \{\psi^{(n)}_L(x) f^{(n)}_L(y)+\psi^{(n)}_R(x) S_n (y) \} \\
           \psi'_R(x) f_R(y)+\sum_{n=1}^\infty\{\psi'^{(n)}_R(x) f^{(n)}_R(y)+\psi'^{(n)}_L(x) S_n (y) \}  \\
\end{array}
\right],\eea where $\psi_{L,R}=(\psi_{1}\ \psi_{2})^T_{L,R}$ and
$\psi'=\psi_3$. Denoting $M_{\psi n}=\sqrt{M^2_\psi+M^2_n}$, the
mode functions are written as $S_n(y)=\fr{1}{\sqrt{\pi
R}}\sin(M_ny)$ and \bea  f_L(y)&=& \sqrt{\fr{M_\psi}{1-e^{-2\pi R
M_\psi}}}e^{-M_\psi |y|},\hs f_R(y)=
\sqrt{\fr{M_\psi}{e^{2\pi R M_\psi}-1}}e^{M_\psi |y|},\label{zmf}\\
f_L^{(n)}(y)&=&\fr{M_n}{\sqrt{\pi R}M_{\psi n}}\left[\cos(M_n y)-\fr{M_\psi}{M_n}\epsilon(y)\sin(M_n y)\right],\\
f_R^{(n)}(y)&=&\fr{M_n}{\sqrt{\pi R}M_{\psi n}}\left[\cos(M_n
y)+\fr{M_\psi}{M_n}\epsilon(y)\sin(M_n y)\right].\eea If $M_\psi$
is positive, the zero mode $f_L$ in (\ref{zmf}) is concentrated at
$y=y_0=0$ while the zero mode $f_R$ is concentrated at
$y=y_\pi=\pi R$. Vice versa, if it is negative the $f_L$ is
concentrated at $y=y_\pi$ while the $f_R$ is concentrated at
$y=y_0$. This can therefore give a very simple realization that
although the Yukawa couplings in the gauge-Higgs unification
original from the gauge coupling, they could be small and
hierarchical (due to difference in the bulk masses), without fine
tuning.

At the boundaries (or branes), the 5D gauge symmetry is broken (by
orbifolding) into \be SU(3)_C \otimes SU(3)_W \otimes U(1)_X
\longrightarrow SU(3)_C \otimes SU(2)_L \otimes U(1)_{T_8}\otimes
U(1)_X,\ee where $T_8$ is the generator of $SU(3)_W$.  On the
other hand, an exact residual symmetry that is never broken is the
electric charge operator. It can be obtained by a combination of
diagonal generators of $SU(3)_W \otimes U(1)_X$ due to electric
charge conservation, which is given by \be Q=T_3 +\fr{1}{\sqrt{3}}
T_8 + X. \ee In this case the second component of $H$ is
electrically neutral. Otherwise, if the first component of $H$
considered is electrically neutral, the coefficient of $T_8$ will
change sign. Both the cases are equivalent, therefore we can
consider the first case. It is also noted that \be
Y=\fr{1}{\sqrt{3}}T_8 + X\label{hcmc}\ee is just hypercharge of
the standard model.

Because of the 5D gauge symmetry all the gauge fields including
the Higgs cannot have any tree level explicit mass term. The Higgs
field in the framework cannot get at this level any nonzero vacuum
expectation value (VEV). The electroweak symmetry can only be
dynamically broken by the Wilson line phase,
$e^{i(\la_a/2)\al^a}\equiv Pe^{ig\oint dyA_5 }$, through Hosotani
mechanics. We therefore suppose that \be \langle A_5 \rangle =
\left[ \begin{array}{ccc}
            0 & 0 & 0 \\
            0 & 0 & -i\fr{\al}{4\pi g R}\\
            0 & i\fr{\al}{4\pi g R} & 0\\
\end{array}\right],\ee where $A_5$ is arranged to develop VEV in the direction of $A^7_5$
(in the matrix we have put $\al^7=\al$ for a simplicity). Notice
that the zero mode field has not been normalized yet; otherwise,
$g$ should be replaced by that of 4D and the equation remaining
unchanged. In the following, the usages of such similar ones
should be flexibly understood.

With this VEV, in the neutral gauge sector the $B_\mu$ remains
massless and does not mix with $A^3_\mu$ and $A^8_\mu$. It is
natural to identify the $B_\mu$ as a new physical gauge boson but
encountered with the inconsistent zero-mass (because any mixing
between $B_\mu$ and the photon in this case is unreasonable). In
addition, we can verify that $s^2_W=3/4$ and the electromagnetic
current are not correct, which happen in the same with the
simplest gauge-Higgs unification version. Let us next deal with
such issues. A suggestion is that the $A^8_\mu$ has to be directly
mixed with $B_\mu$ by some source so that the resulting gauge
boson associated with the hypercharge is correctly identified
(instead of $A^8_\mu$), and the resulting new gauge boson (instead
of $B_\mu$) can get heavy. This is possible if we introduce a
heavy scalar field charged under both the $U(1)_X$ and $SU(3)_W$,
so that the VEV of this scalar will at least break
$U(1)_{T_8}\otimes U(1)_X$ symmetry into $U(1)_Y$ of the standard
model and simultaneously providing the mass for the new gauge
boson. This breaking probably lies at the same stage with the
orbifolding breaking.

Denoting the quantum numbers by $(SU(3)_C,SU(3)_W,U(1)_X)$, the
scalar transforms as \be   \chi=\left[ \begin{array}{c}
               \chi^+ \\
               \chi^0 \\
               \chi'^0\\
\end{array}\right]\sim (1,3,1/3),\ee satisfying the boundary condition
\be \chi(x,y_i-y)=P_i \chi(x,y_i+y).\ee The parity is explicitly
written as \be \chi=\left[ \begin{array}{c}
               (-\ -) \\
               (-\ -) \\
               (+\ +)\\
\end{array}\right].\ee Due to parity conservation, only the third component can develop VEV,
which satisfies our minimal requirement for the symmetry breaking,
\be\langle \chi\rangle =\left[ \begin{array}{c}
               0 \\
               0 \\
               \fr{\beta}{4\pi gR}\\
\end{array}\right],\ee with $\beta \gg \al$, i.e. $k\equiv \fr{\beta}{\al}\gg
1$. Here, notice that this VEV always conserves the residual
standard model symmetry. There exist two simultaneous symmetry
breaking processes of the same stage: the symmetry breaking of
$SU(3)_W\rightarrow SU(2)_L$ is a result of orbifolding
characterized by $1/R$ scale and the symmetry breaking of
$U(1)_{T_8}\otimes U(1)_X \rightarrow U(1)_Y$ is due to $\beta$.
The VEV $\beta$ can be naturally taken in the same order with
$1/R$, i.e. $\beta\sim \mathcal{O}(1)$ and thus $\alpha$ close to
zero (this $\al$ should be provided from the effective potential
for the Higgs field). This is dynamical because as any ordinary
scalar field theory the brane field $\chi'^0$ is unstable under
radiative corrections. The VEV $\beta$ that is obtained from the
resulting effective potential for $\chi$ can get naturally
generated in the cut-off scale $1/R$; and the $\chi'^0$ has also a
heavy mass in this scale. With the $\chi$ scalar, all the issues
as stated above are solved (shown explicitly below). In addition,
to make a consistent fermion spectrum the other brane scalars will
be introduced, that provide large masses for exotic fermions such
as right-handed neutrinos and extra chiral leptons and quarks. The
VEVs of these scalars like $\beta$ have similar contributions to
the gauge boson spectrum as shown in the last section
(\ref{fect}), therefore in the following we consider only the case
with the $\chi$ scalar and its VEV $\beta$.

A summary on the symmetry breaking is as follows. The first stage
of symmetry breaking down to that of the standard model is due to
the orbifold boundary conditions and the $\chi$ scalar VEV
$\beta$. The second stage of symmetry breaking from the standard
model symmetry down to that of QCD and QED is due to the Wilson
line phase characterized by $\al$. It is \bea SU(3)_C \otimes
SU(3)_W \otimes U(1)_X \stackrel{1/R,\beta}{\longrightarrow}
SU(3)_C \otimes SU(2)_L \otimes U(1)_Y
\stackrel{\al}{\longrightarrow} SU(3)_C\otimes U(1)_Q, \eea where
the gauge hierarchy is \be \fr{\beta}{R}\sim \fr{1}{R}\gg
\fr{\al}{R}.\ee

\section{\label{gabo}Gauge bosons}
The mass Lagrangian for the zero mode gauge bosons is  \bea
{\mathcal{L}}^{\mathrm{gauge}}_{\mathrm{mass}} &=& \int_{-\pi
R}^{\pi R} dy \left\{g^2{\mathrm{Tr}} [A_\mu,\langle A_5\rangle
]^2 +(D_\mu \langle \chi\rangle)^{\dagger} (D^\mu \langle \chi
\rangle)\right\}\crn  &=& \fr 1 2 \left(\fr{\al}{4\pi
R}\right)^2W^- W^+ + \fr 1 4 \left(\fr{\al}{4\pi R}\right)^2
(A^3-\sqrt{3}A^8)^2+\fr 1 3 \left(\fr{\beta}{4\pi
R}\right)^2(A^8-\fr{t}{\sqrt{3}}B)^2,\label{dd3} \eea where
$t\equiv \fr{g_x}{g}$ and $W^{\pm}\equiv \fr{A^1\mp i
A^2}{\sqrt{2}}$. From the first term, we obtain the mass of $W$
boson: \be m_W=\fr{\al}{4\sqrt{2}\pi R}.\label{mw}\ee The
remaining terms provide the mass Lagrangian for the neutral gauge
bosons that can be rewritten as $\fr{1}{2}(A^3\ A^8\ B)M^2 (A^3\
A^8\ B)^T$, with \bea M^2=\left(\fr{\al}{4\pi R}\right)^2
\left[\begin{array}{ccc}
     \fr 1 2  & -\fr{\sqrt{3}}{2} & 0 \\
     -\fr{\sqrt{3}}{2} & \fr 3 2 +\fr 2 3 k^2 & -\fr{2t}{3\sqrt{3}}k^2 \\
      0 & -\fr{2t}{3\sqrt{3}}k^2 & \fr{2 t^2}{9}k^2\\
\end{array}\right].\label{ngm}\eea
The straightforward procedure for diagonalizing this mass matrix
as well as identifying the Weinberg angle and physical gauge
bosons can be found in \cite{dl}. In details, one can check that
the mass matrix (\ref{ngm}) always has a non-degenerate
zero-eigenvalue, $\mathrm{det} M^2=0$, that corresponds to the
photon. The photon field as associated with the electric charge
operator $Q$ is the corresponding eigenstate obtained by \be
A_\gamma= \fr{\sqrt{3}t}{\sqrt{3+4 t^2}} A^3 +  \fr{t}{\sqrt{3+4
t^2}} A^8 +\fr{\sqrt{3}}{\sqrt{3+4 t^2}}B,\label{photon1}\ee which
is independent of the VEVs. All these are natural consequences of
the electromagnetic gauge invariance \cite{dl}.

Now, with the help of (\ref{photon1}) we can obtain
electromagnetic interactions, e.g choosing a lepton triplet $\psi
= (\nu\ e\ e')^T\sim (1,3,-2/3)$ will yield matching condition of
the gauge coupling constants $e=g s_W$, with \be s_W\equiv
\fr{\sqrt{3}t}{\sqrt{3+4t^2}}\label{sinw}\ee that defines the
Weinberg angle. We can evaluate $t=g_x/g$ so that $s_W$ gets the
correct value. Namely, $t=\sqrt{3}s_W/\sqrt{3-4s^2_W}\simeq 0.58$
provided that $s^2_W\simeq 0.231$ \cite{pdg}. The photon field
(\ref{photon1}) can be rewritten as \be A_\gamma = s_W A^3 + c_W
\left(\fr{t_W}{\sqrt{3}}A^8 +\sqrt{1-\fr{t^2_W}{3}}B\right). \ee
The standard model $Z$ boson is orthogonal to $A_\gamma$ as usual:
\be Z = c_W A^3 - s_W \left(\fr{t_W}{\sqrt{3}}A^8
+\sqrt{1-\fr{t^2_W}{3}}B\right).\ee Notice that the one in
parentheses is just ordinary gauge field as associated with the
hypercharge $Y$ given above. It is a mixing of $A^8$ and $B$.  A
field that is orthogonal to it, i.e. to $A_\gamma$ and $Z$, will
be a new gauge boson: \be Z'=\sqrt{1-\fr{t^2_W}{3}}A^8
-\fr{t_W}{\sqrt{3}}B. \ee

In the new basis $(A_\gamma\ Z\ Z')$, the mass matrix (\ref{ngm})
becomes \be M^2\longrightarrow M'^2= \left[\begin{array}{ccc}
0 & 0 & 0 \\
0 &m^2_Z & m^2_{ZZ'}\\
0& m^2_{ZZ'}&m^2_{Z'}\\
\end{array}\right],\ee where
\bea m^2_Z &=& \fr{3+4 t^2}{2(3+t^2)}\left(\fr{\al}{4\pi
R}\right)^2, \hs m^2_{Z'}= \fr{81+4(3+t^2)^2
k^2}{18(3+t^2)}\left(\fr{\al}{4\pi R}\right)^2,\label{mz}\\
m^2_{ZZ'}&=&-\fr{3\sqrt{3+4t^2}}{2(3+t^2)}\left(\fr{\al}{4\pi
R}\right)^2,\label{mzpp}\eea that defines the mixing of $Z$ and
$Z'$. The mixing angle of $Z-Z'$ is given by \be
\tan(2\phi)=\fr{2m^2_{ZZ'}}{m^2_{Z'}-m^2_Z}.\ee We thus obtain the
following physical gauge bosons \be \mathcal{Z}=c_\phi Z - s_\phi
Z',\hs \hs \mathcal{Z}'=s_\phi Z+ c_\phi Z',\ee with masses\bea
m^2_{\mathcal{Z}}&=&\fr 1 2
(m^2_Z+m^2_{Z'}-\sqrt{(m^2_Z-m^2_{Z'})^2+4
m^4_{ZZ'}}),\label{mz1}\\ m^2_{\mathcal{Z}'}&=&\fr 1 2
(m^2_Z+m^2_{Z'}+\sqrt{(m^2_Z-m^2_{Z'})^2+4 m^4_{ZZ'}}).\eea

Because of $k=\beta/\al\gg 1$, from (\ref{mz}) and (\ref{mzpp}) we
have $m_{Z'}\gg m_Z,\ m_{ZZ'}$. This implies that  the $\mathcal
{Z}'$ is heavy and the mixing angle $\phi$ is small. In details,
the approximations can be calculated as follows:  \bea \phi \simeq
-\fr{(3-4s^2_W)^{3/2}}{4c^4_W}\left(\fr{\al}{\beta}\right)^2\simeq
-\left(\fr{\al}{\beta}\right)^2,\hs\hs m_{\mathcal{Z}'}\simeq
\fr{\sqrt{2}c_W}{\sqrt{3-4s^2_W}}\fr{\beta}{4\pi R}\simeq 0.068
\times \fr{\beta}{R}.\eea Consequently, $\mathcal{Z}\simeq Z$ is
the standard model $Z$ like boson, and $\mathcal{Z}'\simeq Z'$
being a new gauge boson. We have also \be m^2_{\mathcal{Z}}\simeq
m^2_Z=\fr{m^2_W}{c^2_W},\ee which can see from (\ref{mz1}),
(\ref{mz}), (\ref{sinw}) and (\ref{mw}). Strictly, we evaluate the
tree level $\rho$ parameter: \be \rho=\fr{m^2_W}{c^2_W
m^2_{\mathcal{Z}}}\simeq
1+\fr{(3-4s^2_W)(13-16s^2_W)}{16c^4_W}\left(\fr{\al}{\beta}\right)^2\simeq
1+ 2\left(\fr{\al}{\beta}\right)^2,\ee which is absolutely close
to one since $\al\ll \beta$, in good agreement with the data
\cite{pdg}. Anyway, the $\rho$ modifies the standard model
expressions for observables by $m_{\mathcal{Z}}\rightarrow
m_{Z}/\sqrt{\rho}$, $\Ga_{\mathcal{Z}}\rightarrow \rho \Ga_Z$, and
$\mathcal{L}_{\mathcal{Z}}\rightarrow \rho \mathcal{L}_Z$, where
$\mathcal{L}_{\mathcal{Z}}$ is an effective four-fermion neutral
current operator. Detailed analyses can be found in, for example,
\cite{them}. Furthermore, from the global fit \cite{pdg} we have
$1.0001< \rho < 1.0025$, thus $0.7\times 10^{-2} <
\fr{\al}{\beta}<3.5\times 10^{-2}$. Combined with (\ref{mw}) and
taking $\beta=1$, we derive the range of Kaluza-Klein scale: \be
40\ \mathrm{TeV} < \fr{1}{R} < 200\ \mathrm{TeV}.\ee  It is also
easy to derive the range of the $Z-Z'$ mixing angle and
$\mathcal{Z}'$ mass:\bea -12.5\times 10^{-4} < &\phi& < -0.5\times
10^{-4},\\ 2.7\ \mathrm{TeV}<&m_{\mathcal{Z'}}&<13.6\
\mathrm{TeV}.\eea

In summary, $\mathcal{Z}'$ is the heavy neutral gauge boson with
mass in the TeV range. This zero mode, including $\chi'^0$ and all
the excitations of the theory, can be integrated out. The physics
below TeV scale, which is localized at the branes,  is the
standard model symmetry, the ordinary gauge bosons and Higgs
doublet, with a perfect consistency of the Weinberg angle, of the
$\rho$ parameter and of all the currents including electromagnetic
current as in our useful standard model \cite{pdg}. To confirm the
last points, we first notice that $W^\pm=(A^1\mp i A^2)/\sqrt{2}$
and the mixing matrix of the neutral gauge bosons (neglect the
$Z-Z'$ mixing): \bea \left[
\begin{array}{c}
        A^3 \\ \\
        A^8 \\ \\
        B\\
\end{array}\right]=
\left[ \begin{array}{ccc}
      s_W & c_W & 0   \\ \\
         \fr{s_W}{\sqrt{3}} & -\fr{s_W t_W}{\sqrt{3}}& \sqrt{1-\fr{t^2_W}{3}}\\ \\
        c_W \sqrt{1-\fr{t^2_W}{3}}&-s_W \sqrt{1-\fr{t^2_W}{3}}& -\fr{t_W}{\sqrt{3}}\\
\end{array}\right]
     \left[ \begin{array}{c}
        A_\gamma \\ \\
        Z \\ \\
        Z'\\
\end{array}\right].
\eea Substituting them into the covariant derivative with notation
that $e=gs_W$, $g_x=gt=g\sqrt{3}s_W/\sqrt{3-4s^2_W}$ and the form
of the electric charge operator, we get the desirable result: \bea
D_\mu &\supset& \pa_\mu + ig (T_1 A^1_\mu + T_2 A^2_\mu) + ig
(T_3A^3_\mu  + T_8A^8_\mu  + t X B_\mu)\crn &\supset& \pa_\mu+
\fr{ig}{\sqrt{2}}(T^+ W^+_\mu + T^- W^-_\mu)+ ie Q A_{\ga \mu} +
\fr{ig}{c_W}(T_3-s^2_W Q) Z_\mu,\eea where $T^\pm=T_1\pm i T_2$ is
the weak isospin raising or lowering operator.

\section{\label{fect}Fermions}

For each lepton family and each quark family, we introduce two 5D
multiplets: \bea \psi^1_a &=&    \left[\begin{array}{c}
           \nu^1_a \\
           e^1_a \\
           e'_a\\
\end{array}\right]\sim (1,3,-2/3),\hs \hs \psi^2_a=
        \left[\begin{array}{c}
           e^2_a \\
           -\nu^2_a \\
           \nu'_a\\
        \end{array}\right]\sim (1,3^*,-1/3),\\
         Q^1_a &=&    \left[\begin{array}{c}
           u^1_a \\
           d^1_a \\
           d'_a\\
        \end{array}\right]\sim (3,3,0),\hs \hs Q^2_a=
        \left[\begin{array}{c}
           d^2_a \\
           -u^2_a \\
           u'_a\\
        \end{array}\right]\sim (3,3^*,1/3),\eea
where $a=1,2,3$ is family index. With the parity as given before,
the zero modes are \bea \psi_a^1 &=&    \left[\begin{array}{c}
           \nu^1_{aL} \\
           e^1_{aL} \\
        \end{array}\right]\oplus e_{aR} ,\hs \hs \psi_a^2=
        \left[\begin{array}{c}
           e^2_{aL} \\
           -\nu^2_{aL}\\
        \end{array}\right]\oplus  \nu_{aR},\\
        Q_a^1 &=&    \left[\begin{array}{c}
           u^1_{aL} \\
           d^1_{aL} \\
        \end{array}\right]\oplus d_{aR} ,\hs \hs Q_a^2=
        \left[\begin{array}{c}
           d^2_{aL} \\
           -u^2_{aL}\\
        \end{array}\right]\oplus  u_{aR},
        \eea under the standard model symmetry at the branes.

We thus have the desirable lepton and quark singlets $e_{aR}$,
$\nu_{aR}$, $d_{aR}$ and $u_{aR}$. But this rises to the unwanted
light lepton and quark doublets (three for leptons and three for
quarks). Such doublets have to be removed by some mechanism. As in
the literature, we can integrate them out by coupling them to
heavy chiral fermions localized at the same brane they
concentrate: $(N_{aR},\ E_{aR})^T\sim (1,2,-1/2)$ and $(U_{aR},\
D_{aR})^T\sim (3,2,1/6)$ under the standard model symmetry. Here
the quantum numbers are given by $(SU(3)_C,SU(2)_L,U(1)_Y)$. The
4D bare couplings are therefore: \bea &&- (\bar{N}_{aR},\
\bar{E}_{aR})
        \left\{m^{l1}_{ab}\left[\begin{array}{c}
          \nu^1_{bL} \\
          e^1_{bL} \\
       \end{array}\right] + m^{l2}_{ab}  \left[\begin{array}{c}
          \nu^2_{bL} \\
          e^2_{bL} \\
       \end{array}\right]\right\}\crn
       &&- (\bar{U}_{aR},\ \bar{D}_{aR})
        \left\{m^{q1}_{ab}\left[\begin{array}{c}
          u^1_{bL} \\
          d^1_{bL} \\
       \end{array}\right] + m^{q2}_{ab}  \left[\begin{array}{c}
          u^2_{bL} \\
          d^2_{bL} \\
       \end{array}\right]\right\} \crn &&+h.c.,\label{dd1} \eea
where the mass parameters $m^{l1,2}_{ab}$ and $m^{q1,2}_{ab}$ are
in general not diagonal in $a$ and $b$ which could lead to flavor
changing profiles for both the lepton and quark sectors despite
the fact that the theory is original from the gauge principle. The
heavy doublets integrated away are exact the combinations as
appearing in the above parentheses; the remaining ones orthogonal
to them are just the expected doublets of the standard model:
       \bea \psi_{aL} &=&    \left[\begin{array}{c}
           \nu_{aL} \\
           e_{aL} \\
        \end{array}\right] ,\hs \hs Q_{aL}=
        \left[\begin{array}{c}
           u_{aL} \\
           d_{aL}\\
        \end{array}\right],
       \eea
where $\psi^{1,2}_{aL}=U^{1,2}_{ab}\psi_{bL}+\cdots$ and
$Q^{1,2}_{aL}=V^{1,2}_{ab}Q_{bL}+\cdots$ For a detailed analysis,
see \cite{overlap-mixing}.

Now, the Yukawa interactions come from \bea &\int_{-\pi R}^{\pi R}
dy& [\bar{\psi}^1_a i\Ga^5 (igA_5) \psi^1_a + \bar{\psi}^2_a
i\Ga^5 (-ig A^*_5) \psi^2_a+ \bar{Q}^1_a i\Ga^5 (igA_5) Q^1_a +
\bar{Q}^2_a i\Ga^5 (-ig A^*_5) Q^2_a]\crn &\supset& -g O^e_{aLR}
(\bar{\nu}^1_{aL},\ \bar{e}^1_{aL})iH e_{aR} + g O^\nu_{aLR}
(\bar{e}^2_{aL},\ -\bar{\nu}^2_{aL})iH^*\nu_{aR} \crn &&-g
O^d_{aLR} (\bar{u}^1_{aL},\ \bar{d}^1_{aL})iH d_{aR} + g O^u_{aLR}
(\bar{d}^2_{aL},\ -\bar{u}^2_{aL})iH^*u_{aR}\crn &&+h.c.\eea where
$H=(1/2)(A^4_5 -i A^5_5,\ A^6_5-iA^7_5)^T$ and the integrals of
wavefunction overlaps given by
\bea O^e_{aLR} &=& \int^{\pi R}_{-\pi R} dy f_L(\psi^1_a)f_R(\psi^1_a)
\simeq 2\pi R M_{\psi^1_a} e^{-\pi R M_{\psi^1_a}},\\
O^\nu_{aLR} &=& \int^{\pi R}_{-\pi R} dy f_L(\psi^2_a)f_R(\psi^2_a)
\simeq 2\pi R M_{\psi^2_a} e^{-\pi R M_{\psi^2_a}},\\
O^d_{aLR}&=&\int^{\pi R}_{-\pi R} dy f_L(Q^1_a)f_R(Q^1_a)\simeq
2\pi R M_{Q^1_a} e^{-\pi R M_{Q^1_a}},\\ O^u_{aLR}&=&\int^{\pi
R}_{-\pi R} dy f_L(Q^2_a)f_R(Q^2_a)\simeq 2\pi R M_{Q^2_a} e^{-\pi
R M_{Q^2_a}}, \eea  provided that $R M_{\psi, Q}$ is around or
larger than 1. Noting that $\langle H \rangle =(0,\
-i\fr{\al}{4\pi g R})^T$, we obtain the following mass terms \be
-m^e_{ab} \bar{e}_{aL} e_{bR}-m^\nu_{ab} \bar{\nu}_{aL} \nu_{bR}
-m^d_{ab} \bar{d}_{aL} d_{bR}-m^u_{ab} \bar{u}_{aL} u_{bR}
+h.c.\ee where \bea m^e_{ab}&=&\sqrt{2}m_W U^{1*}_{ba}
O^e_{bLR},\hs \hs m^\nu_{ab}=-\sqrt{2}m_W
U^{2*}_{ba}O^\nu_{bLR},\\ m^d_{ab} &=&\sqrt{2}m_W V^{1*}_{ba}
O^d_{bLR},\hs \hs m^u_{ab}=-\sqrt{2}m_W V^{2*}_{ba}O^u_{bLR},\eea
which realize hierarchical and small masses due to the exponent
suppressions of the different bulk masses. We see that the mass of
$W$ boson enters because in the framework the Yukawa couplings are
just the gauge coupling $h=g$. At this stage we could have a
consistent quark sector.  However, the neutrinos have only Dirac
masses, the right-handed neutrinos should also be made heavy. This
is possible because the zero mode $\nu_R$ is singlet under the
standard model symmetry that can be self-coupled at the brane
$y=y_\pi$ to perform a mass term: \be -\fr 1 2
m^R_{ab}\bar{\nu}^c_{aR} \nu_{bR} +h.c.\label{dd2}\ee The $\nu_R$
thus have large Majorana masses $m^R_{ab} \sim \beta/R$ (as shown
below). Hence in the model, the masses of the (effective) light
neutrinos are generated via a type I seesaw mechanism \be
m^{\mathrm{eff}}=-m^\nu (m^{R})^{-1} m^{\nu T}\sim
\left(\fr{m^\nu}{10\ \mathrm{MeV}}\right)^2\times \mathrm{eV},\ee
provided $R^{-1}\sim 100\ \mathrm{TeV}$. The neutrino masses are
in sub eV if the Dirac ones are around values of the first
generation quark and lepton masses.

In the following, charges of the extra brane fermions under the
surviving brane $U(1)_{T_8}$ and $U(1)_X$ will respectively be
assigned by $(N_R,\ E_R)^T\sim (0,-1/2)$ and $(U_R,\ D_R)^T\sim
(0,1/6)$ to cancel anomalies. In addition, such charges for the
other chiral fermions are easily obtained that can be founded in
the next section, and the scalar $\chi'^0\sim (T_8,1/3)$. We can
check that the mass terms in (\ref{dd1}) and (\ref{dd2}) cannot be
lift to Yukawa couplings with $\chi'^0$ due to those surviving
gauge symmetries. We therefore introduce other scalars (which must
be singlet under the standard model symmetry): \ben \item
$\eta_1\sim (T_8,-1/6)$ coupled to all terms in (\ref{dd1}), thus
$m^{l1,q1}_{ab}\sim \langle \eta^*_1\rangle$ and
$m^{l2,q2}_{ab}\sim \langle \eta_1 \rangle $, where the $T_8$ is
that of $SU(2)_L$ doublet (as given below). Let us denote $\langle
\eta_1\rangle\equiv \fr{\beta_1}{4 \pi g R}$.
\item $\eta_2\sim (2T_8,2/3)$ coupled to (\ref{dd2}), thus
$m^R_{ab}\sim \langle \eta_2 \rangle\equiv \fr{\beta_2}{4\pi g
R}$. In this case the $T_8$ is that of $SU(2)_L$ singlet.\een
These scalars will also contribute to the mass of the extra gauge
boson and the mixing, as given in the following terms: \be \fr 1 3
\left(\fr{\beta_1/2}{4 \pi
R}\right)^2\left(A^8-\fr{t}{\sqrt{3}}B\right)^2+\fr 1 3
\left(\fr{2\beta_2}{4 \pi
R}\right)^2\left(A^8-\fr{t}{\sqrt{3}}B\right)^2,\label{dd4}\ee
respectively, to be added to the mass Lagrangian (\ref{dd3}). We
see that these contributions (\ref{dd4}) are similar to the last
term in (\ref{dd3}). This is because the charges of $\eta_1$,
$\eta_2$ and $\chi'^0$ by themselves are aligned in the same
direction. Let us denote $k_1\equiv \fr{\beta_1}{2\al}$ and $k_2
\equiv \fr{2 \beta_2}{\alpha}$. The mass matrix of the neutral
gauge bosons (\ref{ngm}) remains unchanged with the replacement:
$k^2\rightarrow k^2+k^2_1+k^2_2$ (or, in the other words
$\beta^2\rightarrow \beta^2+\beta^2_1/4 +4\beta^2_2$). Hence, the
presence of $\eta_{1}$ and $\eta_2$ does not change our
conclusions. Also, the number values retain if we take, for
example, $\beta=\beta_1/2=2\beta_2=1/\sqrt{3}$.

\section{\label{anoeff}Anomaly cancellation}

\subsection{Bulk anomalies are absent}

Because we have introduced the whole fermions transforming on bulk
as vectorlikes under any bulk gauge group, all the anomalies
including the mixed ones are canceled on bulk \cite{anomaly}.

Concretely, the vectorlikes mean that the left and right
components of fermions transform similarly under the gauge groups,
i.e. on bulk, the generators of $\psi_L$ and $\psi_R$ respectively
satisfy $T_{aL}=U^{-1}T_{aR} U$ for some unitary matrix $U$. Here
$T_{a}$ includes the $U(1)_X$ one as well. It deduces that the
following general anomaly vanishes \be
A_{abc}=\mathrm{Tr}[\{T_{aL},T_{bL}\}T_{cL}-\{T_{aR},T_{bR}\}T_{cR}]=0.\ee

\subsection{Brane anomalies are also absent due to exotic chiral fermions}

At the branes, due to the orbifold boundary conditions the
survival (residual) gauge symmetry is now \be SU(3)_C \otimes
SU(2)_L \otimes U(1)_{T_8} \otimes U(1)_X,\label{odbs}\ee as
associated with the standard model gauge bosons and a new gauge
boson $Z'$. For convenience we will work on the basis
(\ref{odbs}). The basis in which the two $U(1)$s above are changed
to the $U(1)_Y$ and the other one orthogonal to $Y$ as coupled to
$Z'$ as mentioned is equivalent.

The zero mode fermions transform under $(SU(3)_C , SU(2)_L ,
U(1)_{T_8} , U(1)_X)$ as follows: $\psi^1_{aL}\sim
(1,2,T_8,-2/3)$, $\psi^2_{aL}\sim (1,2^*,-T_8,-1/3)$,
$Q^1_{aL}\sim (3,2,T_8,0)$, $Q^2_{aL}\sim (3,2^*,-T_8,1/3)$,
$e_{aR}\sim (1,1,T_8,-2/3)$, $\nu_{aR}\sim (1,1,-T_8,-1/3)$,
$d_{aR}\sim (3,1,T_8,0)$, $u_{aR}\sim (3,1,-T_8,1/3)$. Notice that
the charge $T_8$ is the one embedded into the corresponding
$SU(2)_L$ representation. For examples, if it is a $SU(2)_L$
doublet, its $T_8$ is the first $2\times 2$ block diagonal matrix
in the $3\times3$ Gell-Mann matrix $\fr 1 2 \la_8$; if it is a
$SU(2)_L$ singlet, the $T_8$ is the 33 component of $\fr 1 2
\la_8$.

It is to be noted that the above zero mode fields have their
history as being originally born from the corresponding bulk
fields under the maximal 5D gauge symmetry. Hence, the charges of
$U(1)_{T_8}$ and $U(1)_X$ are not arbitrary. Also, their
hypercharges must be constrained by (\ref{hcmc}). But, the status
of the exotic chiral fermions is different because they do not
have any prehistory. They are not controlled by the higher gauge
symmetries, therefore the condition (\ref{hcmc}) with the
component charges resulting from a decomposition is not applied.
The two charges of such $U(1)$s for the exotic chiral fermions are
somewhat arbitrary, however their hypercharges necessarily get the
correct values and given in a similar form as (\ref{hcmc}). This
is the important point to cancel the chiral anomalies on brane.
Namely, let us put $(U_{aR},D_{aR})^T\sim (3,2,0,1/6)$ and
$(N_{aR},E_{aR})^T\sim (1,2,0,-1/2)$. It is easily checked that
all the brane anomalies are removed, when taking into account all
the fermions as mentioned.

Concretely, let us take some examples as follows. $[SU(3)_C]^3$
anomaly: looking at the particles colored such as $u^1_L$,
$d^1_L$, $u^2_L$, $d^2_L$, $u_R$,  $d_R$, $U_R$ and $D_R$ we see
that the number of left chiral components equals to the number of
right chiral ones, therefore the anomaly is canceled out on every
generation. It is well known that the $[SU(2)_L]^3$ anomaly or
mixed anomalies between $SU(2)_L$ and $SU(3)_C$ are always absent.
The gravity anomalies that are potentially troublesome are
$[gravity]^2 U(1)_{T_8}$ and $[gravity]^2U(1)_X$. The first one is
actually canceled because $T_8$ and $-T_8$ always appear in pair
respective to every two fermions of the same chiral kind left or
right. The second one over every fermion generation is as follows:
\bea \sim \sum_{\mathrm{lepton,\ quark,\
chiral-fermion}}(X_L-X_R)&=&2(-2/3)-(-2/3)+2(-1/3)-(-1/3)\crn
&&+3.2(0)-3(0)+3.2.(1/3)-3(1/3)\crn &&-2(-1/2)-3.2.(1/6)=0.\eea It
is no hard to check that $[SU(3)_C]^2U(1)_{T_8}$ and
$[SU(3)_C]^2U(1)_{X}$ anomalies vanish. The latter is since
$2(1/3)-1/3-2(1/6)=0$ for every generation. The
$[SU(2)_L]^2U(1)_{T_8}$ anomaly is also absent. Now we consider
$[SU(2)_L]^2U(1)_X$ anomaly which for each generation is
proportional to \be \sim \sum_{\mathrm{doublets,\
antidoublets}}(X_L-X_R)=-2/3-1/3+3(1/3)-(-1/2)-3(1/6)=0.\ee
Finally, we check the $[U(1)_{T_8}]^2 U(1)_X$ anomaly: \be
\fr{1}{12} .2(-2/ 3)+\fr{1}{12} .2( -1/3)+3.\fr{1}{12} .2
(1/3)-\fr 1 3 (-2/3)-\fr 1 3 (-1/3)-3.\fr 1 3 (1/3)=0.\ee The
remaining anomalies with the $U(1)$s such as
$[U(1)_X]^2U(1)_{T_8}$, $[U(1)_{T_8}]^3$, and $[U(1)_X]^3$ also
vanish for every generation.

It is noteworthy that the presence of the exotic chiral fermions
in the model is more natural because it makes the fermion content
consistent and the model calculable (as such it makes the useless
fermions heavy, provides the realistic fermion flavor mixings, and
makes the model free from all the anomalies to ensure the
consistency of the theory).

\section{\label{concl}Conclusions}

In this work, we have shown that on the view of the
electromagnetic current, the possible minimal bulk gauge symmetry
responsible for the gauge-Higgs unification could be the one based
on $SU(3)_C\times SU(3)_W\times U(1)_X$. A bulk scalar triplet has
been introduced to interpret the natural consequences of the
model. First, the zero mode field of this scalar must be heavy
because as any other 4D scalar theories the mass parameter is
unstable under radiative corrections, it is natural to take this
parameter in the cut-off scale $1/R$. The VEV of this scalar field
is thus in the same order, which with the orbifold boundary
conditions breaks the bulk gauge symmetry into that of the
standard model, providing the new neutral zero-mode gauge boson
$Z'$ with a corresponding large mass. Second, the explicit mixings
among the zero-mode neutral gauge bosons have been achieved.
Thereby, the identifications of the standard model gauge bosons
and the new $Z'$ are obvious. By the electromagnetic coupling, we
have naturally identified the Weinberg angle as in the case of the
ordinary standard model. The correct value of the Weinberg angle
has been obtained. Since the extra scalar fields $\eta_{1,2}$
similar to the zero mode field of the scalar triplet can live in
the branes, the right-handed neutrinos and exotic chiral leptons
and quarks could be made heavy by coupling to these brane scalars.
The contributions of all the scalars as mentioned to the gauge
boson spectrum are the same. Our general conclusions and number
values have thus been obtained with considering just the zero mode
of the scalar triplet.

A minimal bulk fermion content for quarks and leptons that
includes only the triplets and antitriplets has been introduced.
For every fermion generation, we have also assumed the two
brane-localized chiral-fermion doublets. The consequences are (i)
the unwanted light fermions due to the orbifold projection can be
made heavy, the rest is the standard model fermion spectrum plus
the three right-handed neutrinos; (ii) the fermion masses and
flavor mixings could be obtained through the interplay of the bulk
kink mass terms and couplings to the exotic chiral fermions (since
the fermion mass matrices are given as a product of these
contributions); (iii) The theory is free from any anomalies,
despite the fact that the extra $U(1)_X$ is included. On the other
hand, the small masses of the neutrinos have been generated
through the type I seesaw mechanism. A evaluation has shown that
if the observed neutrino masses are in sub eV, their Dirac masses
are around the mass values of the first generation particles of
the standard model.

The model we proposed is very constrained. From the global fit
data on the $\rho$ parameter, we obtain that the compactification
scale is in the range 40 GeV$ < 1/R < $ 200 GeV which strongly
coincides with the previous studies (see, for example, Y. Adachi
{\it et al.} in \cite{overlap-mixing}). The $Z-Z'$ mixing angle
and the mass of $Z'$ are also given as $-12.5\times
10^{-4}<\phi<-0.5\times 10^{-4}$ and 2.7 GeV $< m_{Z'}< $ 13.6
GeV, which are in good agreement with models other than containing
a new neutral gauge boson $Z'$ \cite{pdg,them}.

Finally, also from the experimental data on the $\rho$ parameter
and the mass of $W$ boson, we have obtained the $\al$ parameter
characterizing for electroweak symmetry breaking in the range:
$0.7\times 10^{-2}< \al < 3.5\times 10^{-2}$ (provided that
$\beta=1$). In practice, the value of $\al$ should be compared
with that obtained from the minimization condition of the
effective potential of the model. Unfortunately, in the theories
of gauge-Higgs unification based on the flat space like ours, such
a small value of $\alpha$ is not easy to derive
\cite{hierarchy-ghu}. There have often been two options to enhance
the electroweak symmetry breaking parameter: (i) tune matter
content appropriately, for example, see G. Cacciapaglia {\it et
al.} in \cite{others}; (ii) construct the model based on the
warped spacetime \cite{wothers}. These issues of the current model
are worth exploring to be devoted for future studies.

\section*{Acknowledgments}
The author would like to thank CERN for hospitality and financial
support during his stay where this work was done. The work is also
supported in part by the National Foundation for Science and
Technology Development (NAFOSTED) of Vietnam.

\appendix


\begin{thebibliography}{99}

\bibitem{martin} See, for example, Stephen P. Martin, hep-ph/9709356.

\bibitem{randallsundrum} L. Randall and R. Sundrum, Phys. Rev. Lett. {\bf 83}, 3370 (1999); 4690 (1999).

\bibitem{add-lh-tc} See, for examples,
                    N. Arkani-Hamed, S. Dimopoulos, and G. Dvali, Phys. Lett. B {\bf 429}, 263 (1998);
                    I. Antoniadis, N. Arkani-Hamed, S. Dimopoulos, and G. Dvali, Phys. Lett. B {\bf 436}, 257 (1998);
                    N. Arkani-Hamed, A. G. Cohen, and H. Georgi, Phys. Lett. B {\bf 513}, 232 (2001);
                    N. Arkani-Hamed, A. G. Cohen, E. Katz, and A. E. Nelson, JHEP {\bf 0207}, 034 (2002).

\bibitem{ghu} D. B. Fairlie, Phys. Lett. B {\bf 82}, 97 (1979);
                             J. Phys. G {\bf 5}, L55 (1979);
              N. S. Manton, Nucl. Phys. B {\bf 158}, 141 (1979);
              Y. Hosotani, Phys. Lett. B {\bf 126}, 309 (1983),
                           Phys. Lett. B {\bf 129}, 193 (1983),
                           Annals Phys. {\bf 190}, 233 (1989).

\bibitem{hierarchy-ghu} H. Hatanaka, T. Inami and C. S. Lim, Mod. Phys. Lett. A {\bf 13}, 2601 (1998);
                        I. Antoniadis, K. Benakli and M. Quiros, New J. Phys. {\bf 3}, 20 (2001);
                        G. von Gersdorff, N. Irges and M. Quiros, Nucl. Phys. B {\bf 635}, 127 (2002);
                        R. Contino, Y. Nomura and A. Pomarol, Nucl. Phys. B {\bf 671}, 148 (2003);
                        C. S. Lim, N. Maru and K. Hasegawa, J. Phys. Soc. Jap. {\bf 77}, 074101 (2008);
                        N. Maru and T. Yamashita, Nucl. Phys. B {\bf 754}, 127 (2006);
                        Y. Hosotani, N. Maru, K. Takenaga and T. Yamashita, Prog. Theor. Phys. {\bf 118}, 1053 (2007).

\bibitem{gaugeredu} Y. Kawamura, Prog. Theor. Phys. {\bf 103}, 613 (2000); {\bf 105}, 691 (2001); 999 (2001).

\bibitem{overlap-mixing} G. Burdman and Y. Nomura, Nucl. Phys. B {\bf 656}, 3 (2003);
                         C. A. Scrucca, M. Serone, L. Silvestrini, Nucl. Phys. B {\bf 669}, 128 (2003);
                         Y. Adachi, N. Kurahashi, C. S. Lim, and N. Maru, JHEP {\bf 1011}, 150 (2010); arXiv:1103.5980.

\bibitem{others} C. Csaki, C. Grojean, and H. Murayama, Phys. Rev. D {\bf 67}, 085012 (2003);
                 Y. Hosotani, S. Noda and K. Takenaga, Phys. Lett. B {\bf 607}, 276 (2005);
                 G. Cacciapaglia, C. Csaki and S. C. Park, JHEP {\bf 0603}, 099 (2006);
                 C. S. Lim and N. Maru, Phys. Rev. D {\bf 75}, 115011 (2007);
                 N. Maru and N. Okada, Phys. Rev. D {\bf 77}, 055010 (2008);
                 N. Maru, Mod. Phys. Lett. A {\bf 23}, 2737 (2008);
                 Y. Adachi, C. S. Lim, and N. Maru, Phys. Rev. D {\bf 76}, 075009 (2007);
                                                    Phys. Rev. D {\bf 80}, 055025 (2009);
                 C. S. Lim, N. Maru, and K. Nishiwaki, Phys. Rev. D {\bf 81}, 076006 (2010).

\bibitem{wothers} Y. Hosotani and M. Mabe, Phys. Lett. B {\bf 615}, 257 (2005);
                  K. Agashe, R. Contino and A. Pomarol, Nucl. Phys. B {\bf 719}, 165 (2005);
                  Y. Sakamura and Y. Hosotani, Phys. Lett. B {\bf 645}, 442 (2007);
                  Y. Hosotani and Y. Sakamura, Prog. Theoret. Phys. {\bf 118}, 935 (2007);
                  A. D. Medina, N. R. Shah and C. E. M. Wagner, Phys. Rev. D {\bf 76}, 095010 (2007);
                  Y. Hosotani, K. Oda, T. Ohnuma, Y. Sakamura, Phys. Rev. D {\bf 78}, 096002 (2008);
                  Y. Hosotani, M. Tanaka, and N. Uekusa, Phys. Rev. D {\bf 82},115024 (2010).

\bibitem{chiralfermion} A. Pomarol and M. Quiros, Phys. Lett. B {\bf 438}, 255 (1998);
                        H. Georgi, A. K. Grant, and G. Hailu, Phys. Rev. D {\bf 63}, 064027 (2001).

\bibitem{anomaly} N. Arkani-Hamed, A. G. Cohen, and H. Georgi,  Phys. Lett. B {\bf 516}, 395 (2001);
                  C. A. Scrucca, M. Serone, L. Silvestrini, and F. Zwirner, Phys. Lett. B {\bf 525}, 169 (2002);
                  L. Pilo and A. Riotto, Phys. Lett. B {\bf 546}, 135 (2002);
                  H. D. Kim, J. E. Kim, and H. M. Lee, JHEP {\bf 0206}, 048 (2002);
                  H. M. Lee, JHEP {\bf 0309}, 078 (2003).

\bibitem{dl} P. V. Dong and H. N. Long, Eur. Phys. J. C {\bf 42}, 325 (2005).

\bibitem{pdg} K. Nakamura {\it et al.} (Particle Data Group), J. Phys. G {\bf 37}, 075021 (2010).

\bibitem{them} P. Langacker and M. Luo, Phys. Rev. D {\bf 44}, 817 (1991); {\bf 45}, 278
(1992); Ref. \cite{pdg}, pp. 137-140.


\end{thebibliography}
\end{document}